\documentclass[twocolumn,showpacs,prl,floatfix,superscriptaddress]{revtex4}
\usepackage{graphicx}
\usepackage{bm}
\usepackage {amsmath}
\usepackage {amssymb}
\usepackage{amsopn} 

\newcommand* {\vek}[1]{\bm{#1}}
\newcommand* {\vekc}[1]{\ensuremath{\bm{\mathcal{#1}}}}
\newcommand* {\kk}{\bm{k}}
\newcommand* {\bohrmag}{\mu_\mathrm{B}}
\newcommand {\pd} [2] {\frac{\partial #1}{\partial #2}}
\newcommand {\td} [2] {\frac{d #1}{d #2}}

\begin{document}

\title {Spin precession and alternating spin polarization in
spin-3/2 hole systems}

\author{Dimitrie Culcer}
\affiliation{Advanced Photon Source, Argonne National Laboratory,
Argonne, IL 60439} \affiliation{Department of Physics, Northern
Illinois University, DeKalb, IL 60115}

\author{C. Lechner}
\affiliation {Institut f\"ur Theoretische Physik, Universit\"{a}t
Regensburg, D-93040 Regensburg, Germany}

\author{R. Winkler}
\affiliation{Advanced Photon Source, Argonne National Laboratory,
Argonne, IL 60439} \affiliation{Department of Physics, Northern
Illinois University, DeKalb, IL 60115}

\date{\today}

\begin{abstract}
  The spin density matrix for spin-3/2 hole systems can be
  decomposed into a sequence of multipoles which has important
  higher-order contributions beyond the ones known for electron
  systems [R.~Winkler, Phys.\ Rev.~B \textbf{70}, 125301 (2004)]. We
  show here that the hole spin polarization and the higher-order
  multipoles can precess due to the spin-orbit coupling in the
  valence band, yet in the absence of external or effective magnetic
  fields. Hole spin precession is important in the context of spin
  relaxation and offers the possibility of new device applications.
  We discuss this precession in the context of recent experiments
  and suggest a related experimental setup in which hole spin
  precession gives rise to an alternating spin polarization.
\end{abstract}


\maketitle


Spin electronics is a quickly developing research area that has
yielded considerable new physics and the promise of novel
applications \cite{wol01}. Among the main focuses of spin
electronics are semiconductor systems, where ferromagnetic
semiconductors and spin polarized transport stand out as major
areas of interest. In these fields, the importance of holes as
compared with electrons is manifold. Firstly, the compound
semiconductors exhibiting ferromagnetism are $p$-type materials,
in which ferromagnetism has been shown to be mediated by the
itinerant holes \cite{mac05a}. These materials have also been used
as sources of spin-polarized holes, motivating the search for a
more complete understanding of hole spin precession and
relaxation. Secondly, the spin-Hall effect was first studied in
the context of a hole Hamiltonian \cite{mur03, cul04} and first
observed in a two-dimensional hole gas \cite{wun05}. With these
facts in mind, we devote this Letter to an in-depth study of the
spin dynamics of hole systems.

The conduction band of common semiconductors like GaAs is
described by a spin-1/2 Hamiltonian so that electron spin dynamics
are relatively amenable to theoretical investigation. It is well
known that for spin-1/2 electrons the spin-orbit interaction can
always be written as a wave vector-dependent effective magnetic
field \cite{dya72, pik84}. Electron spin precession in this
effective field, as well as in an external field, has been
discussed extensively and is well understood \cite{win04}. The
spin dynamics of holes, described by an effective spin  $s = 3/2$
\cite{lut56}, have been studied to a lesser extent \cite{jia05,
cul05a}. In general, the coupling of the hole spin and orbital
degrees of freedom cannot be written as an effective magnetic
field, so the simple picture of a spin precessing around an
effective Zeeman field is not applicable to holes. Nevertheless,
we will show that the spin dynamics of hole systems {\it can} be
viewed as a precession, if precession is understood as a
nontrivial {\it periodic} motion in spin space described by an
equation of the type $\td{\mathcal{S}}{t} = \frac{i}{\hbar}
[\mathcal{H}, \mathcal{S}]$ for a suitably generalized spin
operator $\mathcal{S}$ and spin Hamiltonian $\mathcal{H}$. While
it has long been known that holes lose their spin information much
faster than electrons, to our knowledge a quantitative discussion
of hole spin precession based on a firm theoretical footing has
not been attempted in the literature. In particular, we will show
that in appropriate geometries the spin polarization can alternate
in time, even at magnetic field $B=0$. This alternating spin
polarization is a dynamical effect specific to spin-3/2 systems.

The explicit form of the spin-orbit interaction depends
sensitively upon the symmetry of the system \cite{bir74, win03,
shy06}, a fact which is captured in an invariant decomposition of
the spin density matrix \cite{win04b}. Neglecting small terms with
cubic symmetry, the spin density matrix of spin-3/2 systems can be
expressed in terms of invariants of SU(2). These are proportional
to spherical tensors -- a monopole, dipole, quadrupole and
octupole, denoted respectively by $M^0$ (a multiple of the
identity matrix), $\vek{M}^{1}$ (the vector of spin-3/2 matrices),
$\vek{M}^{2}$ and $\vek{M}^{3}$:
\begin{equation}
  \label{eq:rhodecomp}
  {\rho} = \rho_0 \, M^0  + \vekc{S}\cdot\vek{M}^{1}
  + \vekc{Q}\cdot\vek{M}^{2} + \vekc{O}\cdot\vek{M}^{3}.
\end{equation}
The dot product for spherical tensors appearing in Eq.\
(\ref{eq:rhodecomp}) is defined in Ref.\ \cite{edm60}. The moments
$\rho_0$, $\vekc{S}$, $\vekc{Q}$, and $\vekc{O}$ provide a set of
\emph{independent} parameters characterizing the weights of the
multipoles in  the expansion (\ref{eq:rhodecomp}) \cite{win04b}.
Excepting normalization factors, the monopole moment $\rho_0$ is
the carrier density, while the dipole moment $\vekc{S}$
corresponds to the spin polarization (Bloch vector) at $B>0$. The
quadrupole moment $\vekc{Q}$ reflects the splitting between the
heavy holes (HHs, spin-$z$ projection $m_s = \pm 3/2$) and the
light holes (LHs, $ m_s = \pm 1/2$). The octupole moment
$\vekc{O}$ is a unique feature of $s=3/2$ systems at $B>0$. We
note that in the density matrix of spin-1/2 electrons only the
first two terms of the expansion (\ref{eq:rhodecomp}) are present.
We would like to point out that the term ``multipole'' is used
here in a rather different sense from that of Ref.~\cite{cul04}.

The dynamics of spin-$3/2$ hole systems are determined by the
\mbox{$4\times 4$} Luttinger Hamiltonian $\mathcal{H}$
\cite{lut56, lip70}, which, in the spherical approximation, reads
\begin{eqnarray}
\mathcal{H} & = & - \frac{\hbar^2 \gamma_1}{m_0} \mathcal{K}^0 \,
M^0 - 2 \sqrt{5} \bohrmag \kappa \, \vekc{K}^1\cdot\vek{M}^1 +
\sqrt{6}\frac{\hbar^2\bar{\gamma}}{m_0} \vekc{K}^2\cdot{\vek M}^2
\nonumber\\ [1.2ex] & \equiv & \mathcal{H}^0 + \mathcal{H}^1 +
\mathcal{H}^2, \label{eq:Hdecomp}
\end{eqnarray}
where $\gamma_1$, $\bar{\gamma}$, and $\kappa$ are Luttinger
parameters, $m_0$ is the bare electron mass, and $\vekc{K}^{j}$
are tensor operators \cite{win04b}. In the lowest order of $\kk$
and $\vek{B}$ we have $\mathcal{K}^0 = k^2$ and $\vekc{K}^1 =
\vek{B}$. The operator $\vekc{K}^2$ is responsible for the HH--LH
coupling in hole systems. We neglect the terms with cubic
symmetry, which are small corrections in the present context, as
discussed in Refs.~\cite{win04b,lip70}.

We will analyze in detail the Heisenberg equations of motion (HEM)
of the multipoles $\vek{M}^j$. Equation (\ref{eq:Hdecomp})
suggests that we study first the HEM
\begin{equation}
  \label{eq:momentprec}
  \td{{\vek{M}}^j}{t} =
  \frac{1}{i\hbar} \; [\vek{M}^j, \mathcal{H}^{j'}] \, .
\end{equation}
The HEM of a dipole $\vek{M}^1$ evolving under the action of the
Zeeman term $\mathcal{H}^1$ describe the well-known Larmor
precession of holes in a magnetic field \cite{lanIV1e}. These
equations are closed, i.e., the RHS does not depend on the other
moments $\vek{M}^{j''}$, $j'' \ne 1$. In general, however, the HEM
for the $\vek{M}^j$ in spin-$3/2$ hole systems cannot be
decoupled, having important consequences for hole spin precession.
This can be seen from Table~\ref{tab:asymirrdec} which gives the
invariant decomposition of the RHS of Eq.\ (\ref{eq:momentprec})
for different $j$ and $j'$ \cite{bir74}. The most remarkable entry
in the table is the one for the quadrupole $\vek{M}^2$ propagating
in time due to a quadrupole term in the Hamiltonian,
$\mathcal{H}^2$, which represents the spin-orbit interaction that
gives rise to the HH--LH coupling. The HEM for $\vek{M}^2$ are
\emph{not} closed. A quadrupole $\vek{M}^2$ precessing in a
quadrupole field ``decays'' into a dipole $\vek{M}^1$ and an
octupole $\vek{M}^3$. This implies that spin precession of an
initially unpolarized system can give rise to spin polarization,
even though there is no external or effective magnetic field. [As
mentioned above, we use the term spin precession for any HEM
(\ref{eq:momentprec}) with $j, j' > 0$.] Moreover, spin precession
of hole systems does not preserve the magnitude of the Bloch
vector, as will be shown below explicitly. These observations
point to the fact that the spin dynamics of spin-3/2 hole systems
are qualitatively different from those of spin-1/2 electron
systems, where the HEM for the monopole (number density) is
decoupled from that of the dipole (Bloch vector $\vek{\sigma}$).
As a result, electron spin precession preserves the length of the
Bloch vector $\langle \vek{\sigma} \rangle$, i.e., $d |\langle
\vek{\sigma} \rangle| / dt = 0$.

\begin{table}[tbp]
  \caption{\label{tab:asymirrdec}Irreducible representations
  $\mathcal{D}^{J''}$ of $SU(2)$ of the (linear combinations of)
  multipoles $\vek{M}^{ J''}$ contained in an invariant
  decomposition of $(1/i\hbar) \; [\vek{M}^j,
  \mathcal{H}^{j'}]$.}
  $\arraycolsep 1em
   \begin{array}{c@{\hspace{2em}}cccc} \hline\hline
    & \mathcal{H}^0 & \mathcal{H}^1 & \mathcal{H}^2 & \mathcal{H}^3
    \\ \hline
         M^0  & 0 & 0 & 0 & 0 \\
    \vek{M}^1 & 0 & \mathcal{D}^1 & \mathcal{D}^2 & \mathcal{D}^3 \\
    \vek{M}^2 & 0 & \mathcal{D}^2 & \mathcal{D}^1 + \mathcal{D}^3
                  & \mathcal{D}^2 \\
    \vek{M}^3 & 0 & \mathcal{D}^3 & \mathcal{D}^2
                  & \mathcal{D}^1 + \mathcal{D}^3 \\ \hline\hline
  \end{array}$
\end{table}

\begin{figure}[tbp]
 \includegraphics[width=0.4\columnwidth]{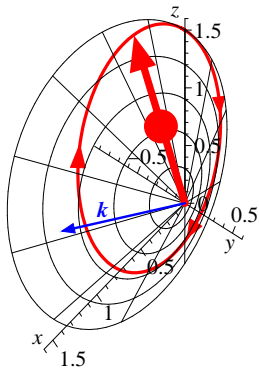}
 \caption{\label{fig:bloch}Precession of the Bloch vector $\vek{S}$
 (bold arrow) about the wave vector $\kk$ when the angle between the
 two is initially $60^\circ$. We have assumed $\vek{S}(t=0)$ is
 parallel to the $z$ axis, and $\kk$ lies in the $xz$ plane.}
\end{figure}

We want to determine and interpret the explicit time evolution of
$\rho$ in order to investigate the physics absent in electron
systems. This calculation is most easily carried out in the
Schr\"odinger picture, which reflects the equivalence of the
Heisenberg and Schr\"odinger pictures for this problem. In the
absence of external fields and disorder, the density matrix
satisfies the quantum Liouville equation $\pd{\rho}{t} =
\frac{i}{\hbar}\,[\rho, \mathcal{H}]$, with formal solution $\rho
(t) = e^{-i\mathcal{H} t/\hbar} \rho (0) e^{i\mathcal{H}
t/\hbar}$, where $e^{i\mathcal{H} t/\hbar}$ is the time evolution
operator (which can often be evaluated in closed form). We
consider first an example where we assume the hole spin to be
oriented initially along the $z$-direction, $m_s = + 3/2$, so that
\begin{equation}
  \label{eq:densini}
  \tilde{\rho} (t=0) = \frac{1}{2} M^0_0  + \frac{3}{2\sqrt{5}} M^1_0
  + \frac{1}{2} M^2_0 + \frac{1}{2\sqrt{5}} M^3_0,
\end{equation}
where the tilde indicates that $\rho$ has been normalized with
respect to the total density $2\rho_0$. This equation demonstrates
that, in general, the density matrix of holes cannot be written
simply as the sum of a monopole and a dipole. The higher
multipoles will be present \cite{win05}. We want to restrict
ourselves to $\mathcal{H} = \mathcal{H}^0 + \mathcal{H}^2$, i.e.,
$B=0$. Table~\ref{tab:asymirrdec} shows that $\rho$ evolves as a
combination of a dipole, a quadrupole, and an octupole. The
implications of this fact for the spin are seen by following the
motion of the Bloch vector $\vek{S}(t) = \frac{3}{2\sqrt{5}} \,
\tilde{\vekc{S}} (t)$
\begin{equation}
  \label{eq:bloch}  \arraycolsep 0.3ex
  \begin{array}{rl}
\vek{S} (t) = \displaystyle \frac{3}{2}\big\{ & \hat{\vek{S}}_0
\left[ \cos^2 (\omega t)
    + c^4 \sin^2 (\omega t) \right] \\ [1.0ex]
  & \displaystyle
  {} + \hat{\kk}\, c\,s\, (1 + c^2)\, \sin^2 (\omega t)
 \\ [1.2ex]
  & \displaystyle
  {} + (\hat{\vek{S}}_0 \times \hat{\kk}) \, 2\, c\, s\,
\sin(\omega t) \cos(\omega t)\big\},
  \end{array}
\end{equation}
where the unit vector $\hat{\vek{S}}_0$ is the orientation of
$\vek{S}$ at $t=0$; $\omega = \bar{\gamma}\hbar k^2/m_0 =
(\varepsilon_h - \varepsilon_l)/2\hbar$, with $\varepsilon_h$ and
$\varepsilon_l$ the HH and LH energies, respectively; $c =
\hat{\vek{S}}_0 \cdot \hat{\kk}$ is the cosine of the angle between
$\hat{\vek{S}}_0$ and $\kk$; and $s$ is the sine of the same angle.
When $\vek{S}_0$ is parallel to $\kk$ (i.e., $c = 1$), we get
$\vek{S} (t) = \vek{S}_0$, which is due to the fact that the initial
state is an eigenstate of the Hamiltonian. In general, neither the
magnitude nor the orientation of the Bloch vector are conserved.
This is illustrated in Fig.~\ref{fig:bloch} showing $\vek{S} (t)$
for an angle of $60^\circ$ between $\hat{\vek{S}}_0$ and $\kk$.
Helicity $\vek{S}\cdot{\kk}$ is conserved, a well-known fact about
this model, which sheds additional light on spin precession in hole
systems. Since $\td{}{t}(\vek{S}\cdot{\kk}) = 0$ and the wave vector
is not changing, $\td{\vek{S} }{t}\cdot {\kk} = 0$. Therefore,
whenever the magnitude of the spin changes, the angle between spin
and wave vector must change in order to preserve the projection of
$\vek{S}$ onto ${\kk}$. As a consequence, no nontrivial spin
precession occurs when $ \vek{S} \perp \kk $ (or $ \vek{S} \parallel
\kk$). We note that energy is also conserved for spin precession in
hole systems. Yet for $B>0$, when $\mathcal{H} = \mathcal{H}^0 +
\mathcal{H}^1 + \mathcal{H}^2$, it can be shown that, in the general
case, energy is transferred back and forth between $\mathcal{H}^1$
and $\mathcal{H}^2$ as time progresses.

\begin{figure}[tbp]
 \includegraphics[width=0.95\columnwidth]{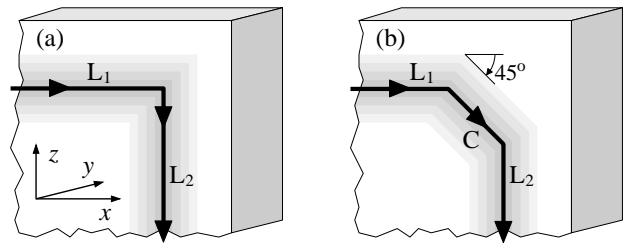}
 \hfill
 \caption{\label{fig:grayson}%
 A bent structure allows holes to be driven around a corner
 \cite{gra04, gra05}.
(a) An idealized representation of the corner device.
(b) A more realistic model of the corner, as discussed in the
text. In regions C and L$_2$ the spin polarization oscillates as a
function of time.}
\end{figure}

We would like to relate our work to recent experiments by Grayson
\emph{et al.} \cite{gra04, gra05} which demonstrated that a
high-quality bent heterostructure can be grown on top of a
pre-cleaved corner substrate that allows one to drive the charge
carriers around an atomically sharp $90^\circ$ corner. Grayson's
experiments were performed on a two-dimensional (2D) electron
system in a GaAs/AlGaAs quantum well. We show here that a similar
system containing holes gives rise to fascinating new physics. The
setup is sketched in Fig.~\ref{fig:grayson}(a). We assume $B=0$ so
that the Hamiltonian is $\mathcal{H} = \mathcal{H}^0 +
\mathcal{H}^2$. An unpolarized HH wave packet travels in the 2D
channel L$_1$ in the $+x$ direction. For simplicity we assume that
all orbital degrees of freedom are integrated over, so that
${\rho}$ depends only on the spin indices and does not depend on
the wave vector $\kk$. Nevertheless, the coupling of spin and
$\kk$ is present in the Hamiltonian through $\mathcal{H}^2$. The
magnitudes of the normalized moments at $t \le0$ (i.e., before
reaching the corner) are $\tilde{\mathcal{S}} =
\tilde{\mathcal{O}} = 0$ and $\tilde{\mathcal{Q}} = 1/2$. The spin
quantization axis of the HH states in L$_1$ is parallel to the
$z$-direction. After the wave packet has passed the corner, the HH
states are not eigenstates of $\mathcal{H}^2$, their spin
quantization axis being perpendicular to the spin quantization
axis supported by the quadrupole field in L$_2$. Therefore, the
quadrupole and octupole moments in L$_2$ oscillate in time
\begin{subequations}
  \label{eq:m90corner}
  \begin{eqnarray}
    |\tilde{\vekc{Q}} (t)|^2 & = & (1/16) + (3/16) \cos^2 (\omega_z t) \\
    |\tilde{\vekc{O}} (t)|^2 & = & (3/16) \sin^2 (\omega_z t),
  \end{eqnarray}
\end{subequations}
with precession frequency $\omega_z = (\varepsilon_h -
\varepsilon_l)/2\hbar \simeq 2\bar{\gamma}\hbar \pi^2/m_0 w^2$. This
frequency can be tailored by varying the width $w$ of the 2D
channel. For GaAs in the spherical approximation we have
$\bar{\gamma} = 2.58$ and we take $w = 10$~nm, yielding a precession
period $2\pi/\omega_z \approx 11$~ps.

For the simplified geometry of Fig.~\ref{fig:grayson}(a) the spin
polarization $\tilde{\mathcal{S}}$ in L$_2$ remains zero, as
required by the conservation of helicity discussed above. However,
assuming a sharp $90^\circ$ corner for an \emph{ideal} 2D system
is certainly an oversimplification, even when the corresponding
\emph{quasi} 2D system has atomically sharp interfaces
\cite{gra04}. A more realistic treatment can be obtained by
modeling the transition region between the channels L$_1$ and
L$_2$ as a sequence of two $45^\circ$ corners as sketched in
Fig.~\ref{fig:grayson}(b). Once again, an unpolarized HH wave
packet travels in channel L$_1$ in the $+x$ direction, the initial
conditions being the same as in the previous example. If the HH
wave packet enters the central region C at $t=0$, we get for the
squared normalized moments in this region
\begin{subequations}
  \label{eq:m45corner}
  \begin{eqnarray}
    \label{eq:m45corner:s}
    |\tilde{\vekc{S}} (t) |^2 & = & (9/80) \; \sin^2 (\omega_z t)\\
    |\tilde{\vekc{Q}} (t) |^2 & = & 1/64 + (15/64) \cos^2 (\omega_z t) \\
    |\tilde{\vekc{O}} (t) |^2 & = & (39/320) \;  \sin^2 (\omega_z t).
  \end{eqnarray}
\end{subequations}
Equation (\ref{eq:m45corner:s}) shows that the initially unpolarized
hole current acquires an alternating spin polarization
$\tilde{\vekc{S}} (t)$ due to spin precession at $B=0$. In Cartesian
coordinates, the Bloch vector in region C reads $\tilde{\vek{S}}(t)
= [0, -\frac{3}{4\sqrt{5}}\, \sin (\omega_z t), 0]$. When the HH
wave packet enters the channel L$_2$ it continues to precess. We get
for the spin polarization
\begin{equation}
  \label{eq:m2x45corner}
  \tilde{\mathcal{S}_y} = \frac{-3}{8\sqrt{5}}
  \left[\cos^2 \left(\frac{\omega_z T}{2} \right)\sin (\omega_z t)
  + 2\sin (\omega_zT) \sin^2 \left( \frac{\omega_zt}{2} \right)\right].
\end{equation}
Here $T$ is the time required to traverse~C which depends on the
length of C and the magnitude of the in-plane wave vector. It
determines the fraction of the spin polarization in L$_2$ that
will be oscillating. If we take the length of C to be of the order
of the channel width, namely $w=10$~nm, and the initial wave
vector $k_F = 0.1$~nm$^{-1}$, we get $T\sim 0.5$~ps. The amplitude
of $\tilde{\mathcal{S}_y}$ in this case is approximately~0.1. We
omit here the qualitatively similar but more complicated
expressions for $\tilde{\vekc{Q}}(t)$ and $\tilde{\vekc{O}}(t)$.
We note also that the approach in Fig.~\ref{fig:grayson}(b) can be
further extended in a transfer-matrix-like approach in order to
describe more complicated geometries.

We expect hole spin precession to be robust against disorder
provided $\omega_z > 1/\tau_s$, a range experiment suggests is
realistic. Hole spin relaxation times $\tau_s$ in the range of
tens of ps \cite{gan02b} or even one ns \cite{rou92a} have been
observed experimentally for GaAs quantum wells, demonstrating that
the spin precession discussed in this paper cannot be neglected in
the context of hole spin dynamics. We expect that the precession
can be observed experimentally using time-resolved optical
techniques such as Kerr or Faraday rotation spectroscopy.

The examples above identify an interesting potential application of
spin-orbit effects in hole systems. By choosing appropriate initial
conditions the spin polarization in the bulk of the system can be
made to oscillate as a function of time. This oscillating spin
polarization is to be contrasted with the Zitterbewegung discussed
lately \cite{sch05a}, in which the carrier \emph{position}
oscillates as a function of time. Unlike conventional transistors,
which work in the diffusive limit, the examples we discuss are valid
in the ballistic limit, a feature shared with the Datta spin
transistor \cite{dat90}. The situation we describe is nevertheless
different from the spin precession in a Datta spin transistor where
the electrons precess not as a function of time but as a function of
position \cite{dat90, win04}.

The above example illustrates the \emph{controlled} hole spin
precession in a confined geometry in a ballistic regime, where the
holes experience an essentially unidirectional quadrupole field.
The situation is qualitatively different in, e.g., bulk material
in a diffusive regime, where the occupied hole states are
characterized by many different orientations of the wave vector
$\kk$. The quadrupole field is thus randomly oriented so that spin
precession in this field must also be taken into account for spin
relaxation, similar to Dyakonov-Perel (DP) spin relaxation in
electron systems. Previously \cite{uen90, fer93}, spin relaxation
of hole systems was discussed mostly using an Elliot-Yafet-type
approach \cite{pik84}, where spin relaxation is caused by momentum
scattering events. DP spin relaxation, on the other hand, implies
that the spin orientation is lost due to spin precession \emph{in
between} the momentum scattering events. Such spin precession for
the case of holes has been described by Averkiev {\it et al.}
\cite{ave02} but only as a result of the spin splitting induced by
bulk and structure inversion asymmetry in 2D systems. On the other
hand, the spin precession described by Eq.~(\ref{eq:bloch}) is
different in nature, being due to the quadrupole field
$\mathcal{H}^2$ so that it is present also in inversion symmetric
systems. Note also that hole spin precession due to
$\mathcal{H}^2$ is typically one or several orders of magnitude
faster than the precession in the effective magnetic field due to
a broken inversion symmetry so that $\mathcal{H}^2$ can yield the
more important contribution to DP hole spin relaxation. (We use
here the term DP spin relaxation in a generalized sense for any
relaxation based on spin precession, even if motional narrowing
\cite{pik84} is not important.) Equation (\ref{eq:bloch}) suggests
that hole DP spin relaxation in bulk systems occurs on time scales
of the order of $2\pi\omega^{-1} = 2\pi (2\bar{\gamma}\hbar
k^2/m_0)^{-1}$. In bulk GaAs, for holes optically excited with
light at $\sim 800$~nm, recent experiments yielded spin relaxation
times of about 110~fs \cite{hil02}, while the simple estimate
above gives $\sim 200$~fs for this case. (This number depends
sensitively on the wave vector of the occupied hole states that
can only be estimated for the experiments in Ref.~\cite{hil02}.)
We remark that, in analogy with spin relaxation in anisotropic
electron systems \cite{ave02, ave99}, the spin relaxation of hole
systems will be characterized by different relaxation times for
each multipole moment in the density matrix.

The research at Argonne National Laboratory was supported by the
US Department of Energy, Office of Science, Office of Basic Energy
Sciences, under Contract No. W-31-109-ENG-38.

\end{document}